\documentclass[12pt]{article}

\pdfoutput=1

\usepackage[a4paper,text={16.8cm,22.4cm}]{geometry}
\usepackage{cite}
\usepackage{amsmath,amssymb,bm,graphicx,color}
\allowdisplaybreaks 
\newcommand\slurp[1]{#1}
\newcommand\sslurp[2]{#1{#2}}
{\catcode`/=\active \expandafter}%
\slurp{\newcommand/}{/\penalty1000\hskip0pt\relax}
{\catcode`:=\active \expandafter}%
\slurp{\newcommand:}{:\penalty1000\hskip0pt\relax}

\newcommand\addspace{\ifcat\nextchar a\spacefactor999. \else.\fi}
{\catcode`\.=\active \expandafter}%
\slurp{\newcommand.}{\futurelet\nextchar\addspace}

\ifx\href\undefined\def\href#1{}\fi
\ifx\texorpdfstring\undefined\def\texorpdfstring#1#2{#1}\fi

\newcommand\myslash{/} \newcommand\mycolon{:}
\newcommand\doi{{\catcode`/=\active \catcode`:=\active \expandafter}\sslurp\realdoi}
{\catcode`/=\active \catcode`:=\active \expandafter}%
\slurp{\newcommand\realdoi[1]{{\let/=\myslash \let:=\mycolon
                               \edef\raw{{http://dx.doi.org/#1}}\expandafter}%
                               \expandafter\href\raw{doi:#1}}}
\newcommand\eprint[2]{{\escapechar-1%
                       \edef\a{\expandafter\string\csname arXiv\endcsname}%
                       \edef\b{\expandafter\string\csname #1\endcsname}%
                       \edef\c{\expandafter\string\csname #2\endcsname}%
                       \edef\d{\noexpand\href{http://arXiv.org/abs/\c}}%
                       \ifx\a\b\expandafter\d\fi{\tt #1:#2}}}

\newcommand{\be}{\begin{equation}}
\newcommand{\ee}{\end{equation}}

\def\d{{\rm d}}
\def\OMIT#1{{}}

\newcommand{\dEM}[1]{\ensuremath{\delta_{\rm EM}^{#1}}}
\newcommand{\dSU}[1]{\ensuremath{\delta_{\rm SU(2)}^{#1}}}
\newcommand{\tdSU}[1]{\ensuremath{\tilde{\delta}_{\rm SU(2)}^{#1}}}

\makeatletter
\DeclareRobustCommand\bfseries{%
  \not@math@alphabet\bfseries\mathbf
  \fontseries\bfdefault\selectfont\boldmath}

\makeatother

\begin{document}

\begin{titlepage}
\begin{flushright}
\end{flushright}

\vspace{0.2cm}
\begin{center}
\LARGE\bf
\bf{Predicting the $\tau$ strange branching ratios \\
$ $
\\
and implications for $V_{us}$}
\end{center}

\vspace{1.cm}
\begin{center}
{\large
Mario Antonelli$^a$, Vincenzo Cirigliano$^{b}$, Alberto Lusiani$^{c}$\\
 and Emilie Passemar$^b$}

\vspace{1.cm}
{\sl 
${}^a$\,  INFN, Laboratori Nazionali di Frascati 
Via E. Fermi 40, I-00044, Italy \\[2mm]
}

\vspace{0.6cm}
{\sl 
${}^b$\, Theoretical Division, Los Alamos National Laboratory  \\
MS B283, Los Alamos, NM 87545, U.S.A.\\[2mm]
}

\vspace{0.6cm}
{\sl 
${}^c$\, INFN Sezione di Pisa;\\
 Scuola Normale Superiore di Pisa, I-56127 Pisa, Italy\\[2mm]
}

\end{center}

\vspace{0.2cm}
\begin{abstract}
\vspace{0.2cm}
\noindent 

Hadronic $\tau$ decays provide several ways to  extract the Cabbibo-Kobashi-Maskawa (CKM)  matrix element $V_{us}$. 
The most precise determination involves using inclusive $\tau$ decays and requires as input the total branching ratio into strange final states.   
Recent results from B-factories have led to a discrepancy of  about $3.4 \sigma$ 
from the value of $V_{us}$ implied by CKM unitarity and direct determination from Kaon semi-leptonic modes. 
In this paper we predict the three leading strange $\tau$ branching ratios, using dispersive parameterizations of the 
hadronic form factors and  taking as experimental input  the measured Kaon decay rates and the $\tau \to K \pi \nu_\tau$ decay spectrum. 
We then use  our results to  reevaluate $V_{us}$, for which we find  $|V_{us}|=0.2207 \pm 0.0027$, in better agreement with 
CKM unitarity. 

\end{abstract}
\vfil

\end{titlepage}


\section{Introduction}

Inclusive hadronic decays of the $\tau$ lepton provide a unique laboratory to study QCD at low energy~\cite{Braaten:1991qm}.
However, predicting exclusive decay rates is a notoriously difficult task, that requires knowing  
the relevant non-perturbative form factors over a wide kinematical range.  
While near threshold rigorous chiral perturbation theory (ChPT) methods can be employed, 
the allowed kinematical region extends well into the resonance domain, 
where  different non-perturbative tools are needed,  such as a combination of dispersion relations and data.

Focusing on $\tau$ decays into strange hadrons  (see Tab.~\ref{Tab:TauBrstrange}, adapted from Ref.~\cite{Amhis:2012bh})  
one notices that   $\Gamma_{10} \equiv \Gamma_{\tau^- \to K^- \nu_\tau}$, 
$\Gamma_{16} \equiv \Gamma_{\tau^- \to K^- \pi^0 \nu_\tau}$ and $\Gamma_{35} \equiv \Gamma_{\tau^- \to \pi^- \bar{K}^0 \nu_\tau}$, 
which represent $68\%$ of the total strange width,  are crossed channels from kaon physics. 
This suggests that,  assuming lepton universality,  one can 
predict $\Gamma_{\tau^- \to K^- \nu_\tau}$, $\Gamma_{\tau^- \to K^- \pi^0 \nu_\tau}$ and $\Gamma_{\tau^- \to \pi^- \bar{K}^0 \nu_\tau}$ 
using the following ingredients: 
(i)  kaon branching ratios (BRs), precisely measured;
(ii)  shape of the $K \pi$ form factors determined by  a combined fit to the $K_{\ell 3}$ decay distribution and 
the $\tau^- \to K \pi \nu_\tau $ invariant mass distribution using a dispersive parametrization for the form factors as presented 
in Refs.~\cite{Bernard:2011ae,Bernard2013}; 
(iii) theoretical input on the electromagnetic and isospin breaking corrections.

The primary  purpose of this work is to  predict the leading strange $\tau$ branching ratios along the lines outlined above. 
We will then use the predicted BRs to update the extraction of  $V_{us}$ from inclusive $\tau$ decays~\cite{Gamiz:2002nu,Gamiz:2004ar} 
and explore how this affects  the $3.4\sigma$ discrepancy with the extractions  of $V_{us}$ 
based on CKM  unitarity and kaon  decays~\cite{Antonelli:2010yf}.


The paper is organized as follows.
In Section~\ref{sect:2} we review the prediction of $\tau \to K \nu_\tau$ from $K_{\mu 2}$. 
In Section~\ref{sect:3} we discuss all the ingredients needed to 
predict $\tau \to K \pi \nu_\tau$ branching ratios in the Standard Model and give our results and error estimates. 
In Section~\ref{sect:4} we work out the implications of the new predicted strange BRs on the inclusive 
extraction of $V_{us}$, and in Section~\ref{sect:5} we give our conclusions.

\begin{table}[h!!]
\begin{center}
\begin{tabular}{ll}
\hline \\[-1.5ex]
\bf{Branching fraction} & {\bf{HFAG Winter 2012 fit}} \\ [2.pt]
\hline \\[-1.5ex]
$\Gamma_{10} = {K^- \nu_\tau}$ & $(0.6955 \pm 0.0096) \cdot 10^{-2}$\\ [2.pt]
$\Gamma_{16} = {K^- \pi^0 \nu_\tau}$ & $(0.4322 \pm 0.0149) \cdot 10^{-2}$\\  [2.pt]
$\Gamma_{23} = K^- 2\pi^0 \nu_\tau$~(ex.~$K^0$) & $(0.0630 \pm 0.0222) \cdot 10^{-2}$\\ [2.pt]
$\Gamma_{28} = K^- 3\pi^0 \nu_\tau$~(ex.~$K^0,\eta$) & $(0.0419 \pm 0.0218) \cdot 10^{-2}$\\ [2.pt]
$\Gamma_{35} = \pi^- \bar{K}^0 \nu_\tau$ & $(0.8206 \pm 0.0182) \cdot 10^{-2}$ \\  [2.pt]
$\Gamma_{40} = \pi^- \bar{K}^0 \pi^0 \nu_\tau$ & $(0.3649 \pm 0.0108) \cdot 10^{-2}$ \\ [2.pt]
$\Gamma_{44} = \pi^- \bar{K}^0 \pi^0 \pi^0 \nu_\tau$ & $(0.0269 \pm 0.0230) \cdot 10^{-2}$  \\ [2.pt]
$\Gamma_{53} = \bar{K}^0 h^- h^- h^+ \nu_\tau$ & $(0.0222 \pm 0.0202) \cdot 10^{-2}$ \\ [2.pt]
$\Gamma_{128} = {K^- \eta \nu_\tau}$ & $ (0.0153 \pm 0.0008) \cdot 10^{-2}$ \\ [2.pt]
$\Gamma_{130} = K^- \pi^0 \eta \nu_\tau$ & $(0.0048 \pm 0.0012) \cdot 10^{-2}$ \\ [2.pt]
$\Gamma_{132} = \pi^- \bar{K}^0 \eta \nu_\tau$ & $(0.0094 \pm 0.0015) \cdot 10^{-2} $ \\ [2.pt]
$\Gamma_{151} = K^- \omega \nu_\tau$ & $(0.0410 \pm 0.0092) \cdot 10^{-2}$ \\ [2.pt]
$\Gamma_{801} = K^- \phi \nu_\tau (\phi \to KK)$ & $(0.0037 \pm 0.0014) \cdot 10^{-2}$ \\ [2.pt]
$\Gamma_{802} = K^- \pi^- \pi^+ \nu_\tau$~(ex.~$K^0,\omega$) & $(0.2923 \pm 0.0068) \cdot 10^{-2}$ \\ [2.pt]
$\Gamma_{803} = K^- \pi^- \pi^+ \pi^0 \nu_\tau$~(ex.~$K^0,\omega,\eta$) & $(0.0411 \pm 0.0143) \cdot 10^{-2}$ \\ [2.pt]
\hline \\[-1.5ex]
$\Gamma_{110} = X_s^- \nu_\tau$ & $(2.8746 \pm 0.0498) \cdot 10^{-2}$\\ [2.pt]
\hline
\end{tabular}
\caption{{\it HFAG Winter 2012 Tau branching fractions to strange final states
~\cite{Amhis:2012bh}.}}
\label{Tab:TauBrstrange}
\end{center}
\end{table}

\section{$\tau \to K \nu_\tau$ from $K_{\mu 2}$ rate in the Standard Model}
\label{sect:2}

Assuming $\tau-\mu$ universality in the charged weak current, the $\tau \to K^- \nu_\tau$ decay rate can be predicted from the $K \to \mu \nu_\mu$ 
decay rate:
\begin{equation}
\mathrm{BR}(\tau \to K \nu_\tau) = \frac{m_\tau^3}{2 m_K m_\mu^2} \frac{S_{\rm EW}^{\tau}}{S_{\rm EW}^{K}}\left(\frac{1-m_K^2/m_\tau^2}{1-m_\mu^2/m_K^2}\right)^2~
\frac{\tau_\tau}{\tau_K}~R_{\rm EM}^{\tau/K}~\mathrm{BR} (K_{\mu 2})~,
\label{eq:BrtauKnu}
\end{equation} 
with $\tau_\tau = 290.6(1.0)$ fs \cite{Beringer:1900zz} and $\tau_K = 12.384(15)$ ns \cite{Antonelli:2010yf} 
the charged $\tau$ and kaon lifetime respectively.
$S_{\rm EW}^{\tau/K}$   represent the short distance electroweak radiative corrections~\cite{Marciano:1993sh, Erler:2002mv}
evaluated at the scale
$\mu = m_\tau$  and $\mu = m_\rho$, respectively. 
%
%
$R_{\rm EM}^{\tau/K}= 1.0090(22)$ \cite{Decker:1993py} the long-distance electromagnetic corrections. 
Using Eq.~(\ref{eq:BrtauKnu}) one finds 
\begin{equation}
\mathrm{BR}(\tau \to K \nu_\tau) = (0.713 \pm 0.003)\times 10^{-2}~.
\label{eq:BRtauKnu}
\end{equation}

\section{$\tau \to K \pi \nu_\tau$ branching ratios in the Standard Model} 
\label{sect:3}

\subsection{Relating $K \to \pi \ell \bar{\nu}_\ell$ 
and  $\tau \to \bar{K} \pi \nu_\tau$  rates}

The decays  $\tau \to K \pi \nu_\tau$ 
and $K \to  \pi \ell \bar{\nu}_{\ell}$  ($\ell = e, \mu$) 
are generated  by the same underlying quark-lepton level  operator 
in the charged current effective  Lagrangian 
(with the replacement $\tau \leftrightarrow \ell$). 
This is true in the Standard Model  (SM)  and in any extension that respects lepton universality. 
Therefore, the hadronic matrix elements for the above two processes 
are related by crossing.   Considering only the SM operator, 
the   $K \to \pi \ell \bar{\nu}_\ell$  amplitude 
involves 
\begin{eqnarray}
\langle \pi(p_\pi) | \bar{s}\gamma_{\mu}u | K(p_K)\rangle &=& 
(p_K+p_\pi)_\mu f^{K \pi}_+(t) + (p_K-p_\pi)_\mu f_-^{K \pi }(t)~,         
\nonumber \\
&=&
\frac{\Delta_{K\pi}}{t} (p_K-  p_\pi)_\mu f_0^{K \pi } (t)  + 
\left[(p_K + p_\pi)_\mu - \frac{\Delta_{K\pi}}{t}(p_K+p_\pi)_\mu \right] f^{K \pi}_{+} (t)~, \ \ \ \ 
\label{eq:elementKl3}
\end{eqnarray}
where $t=(p_K-p_\pi)^2$ and $\Delta_{K\pi}= m_K^2-m_\pi^2$. 
The vector  (scalar)  form factors $f_+(t)$  ($f_0(t)$)  represent  the P-wave  (S-wave) projection 
of the crossed channel matrix element 
$\langle  K \pi  |\bar{s}\gamma^{\mu}u| 0  \rangle$.
The scalar form factor $f_0(t)$ can be expressed in terms of $f_+(t)$ and $f_-(t)$ as 
$f_0(t)= f_+(t) + t/\Delta_{K \pi}  f_-(t)$, and 
by construction, $f_0(0)=f_+(0)$. 
The hadronic matrix element relevant for $\tau \to K \pi \nu_\tau$ 
reads
\begin{equation} 
\langle  \bar{K} (p_K) \pi (p_\pi)  | \bar{s}\gamma_{\mu}u | 0 \rangle = 
-\frac{\Delta_{K\pi}}{s} (p_K+p_\pi)_\mu f_0^{K \pi } (s) -
\left[(p_K-p_\pi)_\mu - \frac{\Delta_{K\pi}}{s}(p_K+p_\pi)_\mu \right] f^{K \pi}_+ (s)~, 
\label{eq:elementTau}
\end{equation}
with in this case $s=(p_K+p_\pi)^2$. 
The decay rates for  $\tau \to K \pi \nu_\tau$ 
and $K \to  \pi \ell \bar{\nu}_{\ell}$  
involve integrals of the form factors over the appropriate phase space. 
The overall normalization, common to both modes  is controlled by $f_{+}^{K\pi} (0)$. 
It is therefore convenient to factor out $f_{+}^{K^0 \pi^-}(0)$, 
denoted $f_{+}(0)$ in the following, in the $K_{\ell3}$ 
and $\tau \to K \pi \nu_\tau$ decay rates. 
The phase space integrals depend then on the normalized form factors,  defined by
\begin{equation}
\bar f_+(s) = \frac{f_+(s)}{f_+(0)},
\ \bar f_0(s) = \frac{f_0(s)}{f_+(0)},
\ \bar f_+(0) = \bar f_{0} (0) = 1~.
\label{eq:Normff}
\end{equation}

With the above definitions for the hadronic form factors, the $K_{\ell 3}$ decay rate reads
\begin{equation} 
\Gamma (K \to \pi \ell \bar{\nu}_\ell [\gamma])
  = \displaystyle \frac{G_F^2 m_K^5 }{192 \pi^3}\,  C_K^2  \, S_{\rm EW}^{K} \, 
 \left(|V_{us}| f_+(0)\right)^2  \, 
I_{K}^\ell \,  \left(1 + \dEM{K \ell} + \dSU{K \pi} \right)^2~. 
\label{eq:GammaKl3} 
\end{equation} 
Here $S_{\rm EW}^{K}$ represents the short distance electroweak radiative corrections~\cite{Marciano:1993sh, Erler:2002mv}
evaluated at the scale $\mu = m_\rho$, 
$C_{K}$ the Clebsch-Gordan coefficients, 
equal to $1$ for $K^0$ and $1/\sqrt{2}$ for $K^-$.
The quantity  $\dEM{K \ell}$ encodes  the channel dependent long-distance 
electromagnetic corrections~\cite{Cirigliano:2001mk,Cirigliano:2004pv}, 
 and  $\dSU{K \pi}$  the correction for strong  isospin breaking. 
It is defined to parameterize  the difference between the $K \to \pi$ 
and $K^0 \to \pi^-$ form factors, so that   $\dSU{K^0 \pi^-} = 0$ 
and   $\dSU{K^+ \pi^0} \neq 0$. 
Finally,  the dimensionless  phase space integral is given by
\begin{equation}
I_{K}^\ell = \int_{m^2_\ell}^{t_{\rm max}}\!\!  dt\,\frac{1}{m_K^8}\,\lambda^{3/2} 
\left(1+\frac{m^2_\ell}{2t}\right) \left(1-\frac{m^2_\ell}{2t}\right)^2 \left(|\bar f_+(t)|^2 + 
\frac{3m^2_\ell \Delta_{K\pi}^2}{(2t + m^2_\ell)\lambda}\,|\bar f_0(t)|^2\right)~, 
\label{eq:IK}
\end{equation}
with $\lambda=[t-(m_K + m_\pi)^2][t-(m_K - m_\pi)^2]$  and $t_{\rm max} = (m_K - m_\pi)^2$. 

The   $\tau \to \bar{K}  \pi \nu_\tau$  decay rate has a structure similar to 
$\Gamma (K \to \pi \ell \bar{\nu}_\ell [\gamma])$. Including electromagnetic and strong isospin breaking 
corrections one has
\begin{equation} 
\Gamma  (\tau \to \bar{K} \pi \nu_\tau [\gamma]) 
 = \displaystyle\frac{G_F^2 m_\tau^5 }{96 \pi^3}\, ~C_K^{2}~S_{\rm EW}^\tau~\left(|V_{us}| f_+(0)\right)^2 
I_{K}^\tau \left(1 + \dEM{K \tau} + \tdSU{K \pi} \right)^2~.
\label{eq:Gamma} 
\end{equation} 
$S_{\rm EW}^{\tau}$ represents the short distance electroweak radiative corrections~\cite{Marciano:1993sh, Erler:2002mv} 
evaluated at the scale $\mu = m_\tau$.  
$C_{K}$ is the Clebsch-Gordan coefficient defined above. 
$\dEM{K\tau}$   is the channel dependent long-distance 
electromagnetic correction  
and  $\tdSU{K \pi}$  the correction for strong  isospin breaking. 
As before,   $\tdSU{K^0 \pi^-} = 0$  and   $\tdSU{K^+ \pi^0} \neq 0$. 
Note that  $\tdSU{K \pi}  \neq \dSU{K \pi}$ because the $K$ and $\tau$  decay rates 
involve integrals of the form factors over very different energy regions. 
Finally, the dimensionless phase space integral , $I_{K}^\tau$ is given by
\begin{equation}
\label{eq:ITau}
I_{K}^\tau   = \ \frac{1}{m_\tau^2}  \int_{s_{K\pi}}^{m_\tau^2}\!\! \frac{ds}{s\sqrt{s}}\, 
\left(1-\frac{s}{m^2_\tau}\right)^2 
\left[\left(1+\frac{2s}{m^2_\tau}\right)~q^3_{K\pi}(s) |\bar f_+(s)|^2+ 
\frac{3 \Delta_{K\pi}^2}{4s}q_{K\pi}(s)\,|\bar f_0(s)|^2\right],  
\end{equation}
with $s_{K\pi}=(m_K+m_\pi)^2$ and $q_{K\pi}$ the kaon momentum in the rest frame of the hadronic system:
\begin{equation}
q_{K\pi}=\frac{1}{2\sqrt{s}} \sqrt{\left(s- s_{K\pi}\right) \left(s-t_{K\pi} \right)}\times \theta \left(s- s_{K\pi}\right),
\qquad
t_{K\pi}=(m_K-m_\pi)^2~.
\end{equation}

Taking the ratios of Eqs.~(\ref{eq:GammaKl3}) and (\ref{eq:Gamma}) and multiplying by the ratio of 
$\tau$ and $K$ lifetimes,  one obtains the following  relation for ${\rm BR}(\tau \to \bar{K} \pi \nu_\tau)$ 
in terms of the crossed channel branching fraction  ${\rm BR} (K \to \pi  \ell \bar{\nu}_\ell)$: 
\begin{equation}
\mathrm{BR}(\tau \to \bar{K} \pi \nu_\tau) =  \frac{2 m_\tau^5}{m_K^5}
 \frac{ S_{\rm EW}^\tau}{S_{\rm EW}^K}  \, 
\frac{I_{K}^\tau}{I_{K}^\ell}  \, \frac{\left(1 + \dEM{K \tau} + \tdSU{K \pi} \right)^2}{\left(1 + \dEM{ K\ell} + \dSU{K\pi} \right)^2} 
\,  \frac{\tau_\tau}{\tau_K}  \   \mathrm{BR} (K \to \pi e \bar{\nu}_e)~,
\label{eq:BrtauKpinu}
\end{equation}
We will use the above formula to predict BR$(\tau \to \bar{K} \pi \nu_\tau)$. 
All the  theoretical and experimental quantities involving $K_{\ell 3}$ decays 
in Eq.~(\ref{eq:BrtauKpinu})  are very accurately known~\cite{Antonelli:2010yf}.
The key new ingredients  are the phase space integrals 
$I_{K}^{\tau}$, that require knowledge of the form factors over a wide energy range, 
 and the electromagnetic and isospin-breaking corrections relevant to the $\tau$ decays, 
$ \dEM{K \tau}$ and $\tdSU{K \pi}$.
In what follows, we describe in detail  the evaluation of these three input quantities. 
Before doing that,  we make the following general observations about our approach: 

\begin{itemize}

\item  
In order to compute  $I_{K}^{\tau}$ (see Eq.~(\ref{eq:ITau})),  
we determine  $\bar{f}_{+,0}^{K^0 \pi^-} (s)$ by  a combined fit to the $K_{\ell 3}$ decay distribution and 
the $\tau^- \to K_S  \pi^-  \nu_\tau $ invariant mass distribution using a dispersive parametrization 
for the form factors~\cite{Bernard:2011ae, Bernard2013}. 

\item  Calculations of $\dEM{K\tau}$ and $\tdSU{K^+ \pi^0} \neq 0$ 
are not as robust  as the corresponding quantities for $K$ decays, 
because a rigorous ChPT analysis  can only be performed in a corner of $\tau$ decay phase space. 
However, we will provide in this paper  first estimates for these quantities. 
In order to estimate the electromagnetic effects we will use a point-like description of pions and kaons, 
neglecting all structure-dependent effects both in loops with virtual photons and Bremsstrahlung amplitudes. 
For the strong isospin breaking effects, we will obtain a rough estimate  by 
using a parameterization of the $s$ dependence of  the form factor 
based on a simple resonance model.   
In both cases we will assign  conservative uncertainties to the results we obtain. 

\end{itemize}
One important  consequence of the above discussion is that we will be able to predict BR$( \tau^- \to \bar{K}^0 \pi^- \nu_\tau)$ 
more accurately than  BR$( \tau^- \to {K}^- \pi^0 \nu_\tau)$, since the latter involves the poorly known  $\tdSU{K^+ \pi^0}$.

%

\subsection{$K\pi$ form factors}
\subsubsection{Parametrization of the form factors}

To compute the phase space integrals, $I_{K}^\ell$, 
one needs  to know  the normalized $K \pi$ form factors, $ \bar f_+(s)$ and $ \bar f_0(s)$  in the 
two energy regions  $m_\ell^2 < s < (m_K - m_\pi)^2$ 
(for $K_{\ell 3}$ decays)   and 
$(m_K + m_\pi)^2 < s < m_\tau^2$  (for $\tau \rightarrow \bar{K} \pi \nu_\tau$).  
To this end, a dispersive representation for the form factors has been introduced in Ref.~\cite{Bernard:2011ae}.  
Here we briefly recall the key ingredients of the two parametrizations used. For more detailed see Ref.\cite{Bernard2013}. 
For the scalar form factor, a dispersion relation with three subtractions is written for ln$\bar f_0(s)$, 
one at the Callan-Treiman point and the other two at zero. This leads to the following representation 
for $\bar f_0(s)$
\begin{eqnarray}
\label{eq:Dispfs}
\bar f_0(s)  = \exp \!\! & & \!\! \left[  \frac{s}{\Delta_{K\pi}} \left( \mathrm{ln}C+
(s-\Delta_{K\pi}) \left( \frac{\mathrm{ln}C}{\Delta_{K\pi}} - \frac{\lambda_0'}{m_\pi^2} \right)
 \right. \right.  \nonumber \\
& & \left. \left. + \frac{\Delta_{K\pi}~s~(s-\Delta_{K\pi})}{\pi} \int_{s_{K\pi}}^{\infty}
\frac{ds'}{s'^2}
\frac{\phi_0(s')}
{(s'-\Delta_{K\pi})(s'-s-i\epsilon)}\right) \right]~. 
\end{eqnarray}
The two subtraction constants a priori unknown, ln$C$ $\equiv$ ln$\bar f_0(\delta_{K\pi})$ and $\lambda_0'$, 
the slope of the form factor (the third one being fixed since $\bar f_0(0) \equiv 1$, see Eq.~(\ref{eq:Normff})), are 
determined from a fit to the data. 
$\phi_0(s)$ represents the phase of the form factor. 
In the low energy region 5$s \le s_{\rm cut}$ we use the $S$-wave $I = 1/2$ $K\pi$ scattering phase 
from Ref.~\cite{Buettiker:2003pp}. For the high-energy region, see discussion below. 

A dispersive representation for the vector form factor $\bar f_+(s)$ is built in a similar 
way~\cite{Pich:2001pj,Boito:2008fq,Boito:2010me,Bernard:2011ae}. 
In this case the three subtractions are performed at $s=0$. Hence the dispersive representation for $\bar f_+(s)$ reads:
\begin{equation}
\bar f_+(s) =  \exp \left[ \lambda_+' \frac{s}{m_\pi^2}+\frac{1}{2} \left( \lambda_+'' - \lambda_+'^2 \right) 
\left( \frac{s}{m_\pi^2} \right)^2 +  \frac{s^3}{\pi} \int_{s_{K\pi}}^{\infty} \frac{ds'}{s'^3} \frac{\phi_+(s')}{(s'-s-i\epsilon)}\right]~.
\label{eq:Dispfv}
\end{equation}
Use has been made of $\bar f_+(0) \equiv 1$ to fix one subtraction constant. 
$\lambda_+'$ and $\lambda_+''$ are the two other subtractions constants corresponding to 
the slope and curvature of the form factor. 
They  are determined from a fit to the data. 
As for the phase of the form factor, $\phi_+(s)$, we 
parameterize it as $\tan \phi_+ (s) =  {\rm Im}  \tilde{f}_+ (s) /  {\rm Re}  \tilde{f}_+ (s)$ 
in terms of  a model for the form factor
$\tilde{f}_+(s)$  
that includes two resonances 
$K^*(892)$ and $K^*(1414)$,  with mixing parameter $\beta$, 
see Refs.~\cite{Jamin:2006tk, Jamin:2008qg, Boito:2008fq, Boito:2010me}:
\begin{equation}
\tilde f_+(s) =  \frac{\tilde m_{K^*}^{2}-\kappa_{K^*} \tilde H_{K\pi}(0)+\beta s}{D(\tilde m_{K^*}, \tilde \Gamma_{K^*})}  
- \frac{\beta s }{D(\tilde m_{K^{*'}}, \tilde \Gamma_{K^{*'}})}~,
\label{eq:Fphase}
\end{equation}
with
\begin{equation}
D(\tilde m_{R}, \tilde \Gamma_{R}) =  \tilde m_R^2-s-\kappa_R~{\rm{Re}}~\tilde H_{K\pi}(s)-i\tilde m_R\tilde\Gamma_{R}(s)~.
\label{eq:Dtilde}
\end{equation} 
In this equation, 
$\tilde m_R$ and $\tilde \Gamma_R$ are model parameters and $\tilde \Gamma_{R}(s)$ and $\kappa_R$ are given by:
\begin{equation}
\tilde \Gamma_{R}(s)= \tilde \Gamma_R \frac{s}{\tilde m_R^2} \frac{\sigma_{K\pi}^3(s)}{\sigma_{K\pi}^3(\tilde m_R^2)}~, \qquad 
\qquad 
\kappa_R = \frac{\tilde{\gamma}_R}{\tilde{m}_R} \frac{192 \pi F_K F_\pi}{( \sigma_{K\pi} (\tilde{m}_R^2))^3}
\end{equation}
with $\sigma_{K\pi}(s)=2q_{K\pi}(s)/\sqrt{s}$. 
 $\tilde H_{K\pi}(s)$ is the $K\pi$ loop function in ChPT~\cite{Jamin:2006tk,Jamin:2008qg}. 
%
%
We emphasize here that $\tilde m_{R}$ and $\tilde \Gamma_R$ are model parameters and do not correspond to the physical 
resonance masses and width. To find them one has to find the pole of Eq.~(\ref{eq:Fphase}) or equivalently 
the zero of Eq.~(\ref{eq:Dtilde}) 
on the second Riemann sheet. 
Note that this model inspired by the Gounaris-Sakourai parametrization~
\cite{Gounaris:1968mw, Beldjoudi:1994hi, Pich:2001pj, Jamin:2006tk, Moussallam:2007qc, Jamin:2008qg, Boito:2008fq, Boito:2010me, Bernard:2011ae} 
is built such that the good properties of analyticity, unitarity and perturbative QCD are fulfilled. 
This model is only valid in the $\tau$ decay region. 
Therefore we will use it for $s \le s_{\mathrm{cut}} \sim m_\tau^2$. 
Hence there will be seven parameters to fit from the data: $\lambda_+'$ and $\lambda_+''$ the slope and the curvature 
of the form factor and the resonance parameters used to model the phase: $m_{K^*}$ and $\Gamma_{K^*}$ the mass and 
decay width of K$^*$(892) and $m_{K^{*'}}$ and $\Gamma_{K^{*'}}$ the mass and decay width of K$^*$(1414) and $\beta$ the mixing 
parameter between the two resonances.

For the high-energy region of the dispersive integrals Eqs.~(\ref{eq:Dispfs},\ref{eq:Dispfv}), 
($s \ge s_{\mathrm{cut}} \sim m_\tau^2$) the phase is unknown and following 
Refs.~\cite{Bernard:2006gy,Bernard:2009zm,Bernard:2011ae,Bernard2013}, 
we take a conservative interval between $0$ and $2\pi$ centered at the asymptotic value of the phase which is $\pi$. 
The use of a three time subtracted dispersion relation reduces the impact 
of our ignorance of the phase at relatively high energies. 
The price to pay is that the correct asymptotic behaviour of the two form factors is subjected 
to a set of sum rules derived in \cite{Bernard:2011ae,Bernard2013}, which is used to constrain our fit parameters.
\subsubsection{Determination of the $K\pi$ form factors from $\tau \rightarrow K \pi \nu_\tau$ Belle data and $K_{\ell3}$ analyses}

We perform a combined fit to the Belle data~\cite{Epifanov:2007rf} as well as the $K_{\ell3}$ data~\cite{Antonelli:2010yf}, 
along the lines described in Ref.~\cite{Bernard:2011ae}. 
We minimize the following quantity:
\begin{eqnarray}
\label{eq:chi2} 
\chi^2 \!\! &=& \!\! \sum_{i} \left( \frac{N_i^{\mathrm{theo}}-N_i^{\mathrm{exp}}}{\sigma_{N_i^{\mathrm{exp}}}} \right)^2  + 
{\mathrm{ln}C - \mathrm{ln}C^{K_{\ell3}} \choose \lambda_+'-\lambda_{+}'^{K_{\ell3}}}^T V^{-1} 
{\mathrm{ln}C - \mathrm{ln}C^{K_{\ell3}} \choose \lambda_+'-\lambda_{+}'^{K_{\ell3}}} \\
& & + \left( \frac{\alpha_{2s}-\alpha_{2s}^{\mathrm{sr}}}{\sigma_{\alpha_{2s}^{\mathrm{sr}}}} \right)^2 
+ \left( \frac{\alpha_{2v}-\alpha_{2v}^{\mathrm{sr}}}{\sigma_{\alpha_{2v}^{\mathrm{sr}}}} \right)^2~,\nonumber 
\end{eqnarray}
where $N_i^{\mathrm{exp}}$ and $\sigma_{N_i^{\mathrm{exp}}}$ are respectively, the experimental number of events 
and the corresponding uncertainty in the $i^{\mathrm{th}}$ bin. The theoretical number of events in a given $i$ bin 
is~\cite{Jamin:2006tk, Jamin:2008qg}
\begin{equation}
N_i^{\mathrm{theo}}= N_{\mathrm{tot}} b_w \frac{1}{\Gamma_{\tau \to K\pi \nu }} \frac{d\Gamma_{\tau \to K\pi \nu}}{d\sqrt{s}}(s_i)~,
\label{eq:Ni}
\end{equation}
with $N_{\mathrm{tot}}$, the total number of events, $b_{w}$ the bin width and $\Gamma_{\tau \to K\pi \nu}$ the total 
decay rate given in Eq.~(\ref{eq:Gamma}). We fit the first 76 points from threshold $s_{K\pi}$ to $s_{\rm fit} \sim 1.51$ where our parametrization is expected 
to be reliable. Note that following Refs.~\cite{Jamin:2008qg,Boito:2008fq,Boito:2010me} we exclude from the fit the points 5, 6 and 7 that 
exhibit a bump which is not present in the preliminary BaBar data~\cite{Paramesvaran:2009ec}. 
We have tested that including these points in the fit amounts to increase the $\chi^2$ from 60/68 to 78/71 
without any significant changes in the values of the parameters, which remain within the error bars. 
The second term of Eq.~(\ref{eq:chi2}) encodes the constraints coming from $K_{\ell3}$ analyses where a dispersive parametrization has been used for the 
form factors~\cite{Bernard:2006gy,Bernard:2009zm}. We are using ln$C^{K_{\ell3}}=0.2004 \pm 0.0091$, 
$\lambda_+'^{K_{\ell3}}=(25.66 \pm 0.41)\times 10^{-3}$ and $\rho(\mathrm{ln}C,\lambda_+')=-0.33$ 
from Ref.~\cite{Antonelli:2010yf}. $V$ represents the covariance matrix.  
In the minimization we also impose the constraints given by the sum rules 
Eqs.~(15) and (18) of Ref.~\cite{Bernard:2011ae,Bernard2013}
\footnote{The constraints from the  other two sum-rules, see Ref.~\cite{Bernard:2011ae,Bernard2013} 
are not imposed in the fit, see Eq.~(\ref{eq:chi2}), 
since they are automatically satisfied due to the large band taken for $\phi_{0,\mathrm{as}}$ 
and $\phi_{+,\mathrm{as}}$.}  
with 
$\alpha_{2s} \equiv \frac{\mathrm{ln}C}{\Delta_{K\pi}} - \frac{\lambda_0'}{m_\pi^2}$, 
$\alpha_{2v} \equiv \lambda_+''- \lambda_+'^2$ and 
\begin{equation}
\alpha_{2s}^{\mathrm{sr}} \equiv \frac{\Delta_{K\pi}}{\pi} 
\int_{s_{K\pi}}^{\infty} \frac{ds'}{s'^2} \frac{\phi_0(s')}{(s'-\Delta_{K\pi})}~,
\label{eq:alphassumrule}
\end{equation}
\begin{equation}
\alpha_{2v}^{\mathrm{sr}} \equiv \frac{2 m_\pi^4}{\pi}  \int_{s_{K\pi}}^{\infty} ds' \frac{\phi_+(s')}{s'^3}~.
\label{eq:alphavsumrule}
\end{equation}

\begin{figure}[t]
\begin{center}
\includegraphics[width=0.8\textwidth]{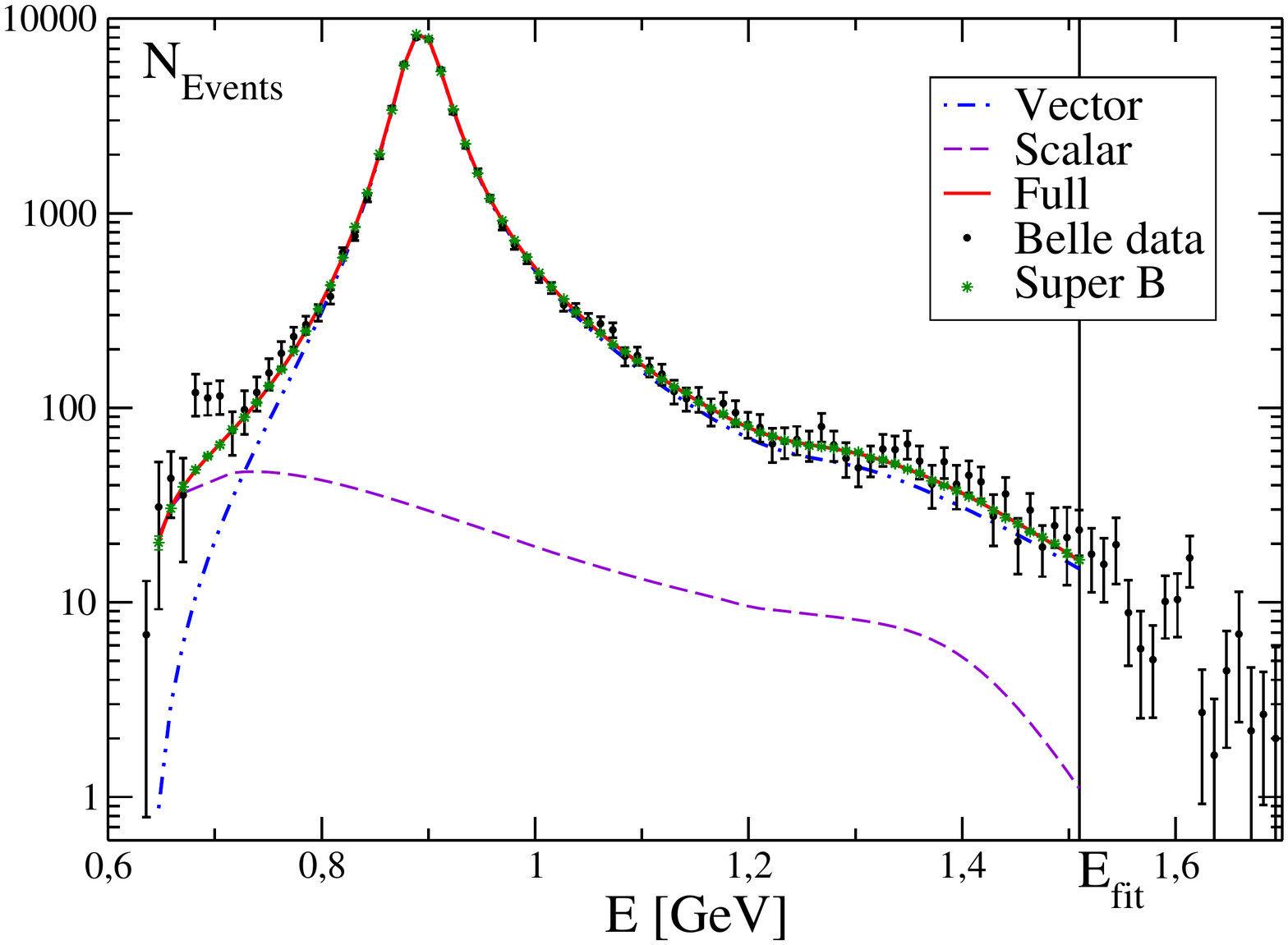}
\caption{\it Fit result for the spectrum of \mbox{$\tau \rightarrow K \pi \nu_\tau$}. 
The data in black are from Belle Collaboration~\cite{Epifanov:2007rf}. The points in green are projected data for 
a second generation B factory with integrated luminosity of $40$~ab$^{-1}$ 
with the same central values of current Belle data and rescaling errors according to the expected sensitivity. 
The dashed violet line represents the scalar form factor contribution. 
The dot-dashed blue line is the vector form factor contribution and the solid red line gives the full result.} 
\label{fig:EventsTaus}
\end{center}
\end{figure}

The results of the fit are presented on Fig.~\ref{fig:EventsTaus} and in Tab.~\ref{tab:fit}
with the correlations between the parameters in  Tab.~\ref{tab:Corrfit}. 
Tab.~\ref{tab:fit} display results for the fit to real data~\cite{Epifanov:2007rf} and also 
projected data from a super-B factory, obtained by keeping the same central values 
of current Belle data~\cite{Epifanov:2007rf} and rescaling the errors according to 
the expected sensitivity of a second generation B factory assuming an integrated luminosity of 
$40$~ab$^{-1}$, see e.g. Ref~\cite{2010arXiv1011.0352A}. 
Using these results we can compute the phase space integrals Eqs.~(\ref{eq:IK}, \ref{eq:ITau})
given in Tabs.~\ref{Tab:IkBelle} and \ref{Tab:IkSuperB}. 


\begin{table}[h!]
\begin{center}
\begin{tabular}{|l||c|c|}
\hline
& $\tau \rightarrow K \pi \nu_\tau~\&~K_{\ell3}$ & $\tau \rightarrow K \pi \nu_\tau ~\&~K_{\ell3}$   \\
& Belle & 2$^{\rm nd}$ generation B factory \\
&  &  (projected) \\
\hline\
ln $C$  & $ 0.20352 \pm 0.00890$ &$ 0.19880  \pm 0.00498$\\
$\lambda_0' \times 10^{3}$ & $13.824 \pm 0.824$& $13.703 \pm 0.521$ \\
\hline
\hline
$\tilde m_{K^*} \mathrm{[MeV]} $  &   $943.59 \pm 0.58 $ & $943.76 \pm 0.06 $ \\
$ \tilde \Gamma_{K^*} \mathrm{[MeV]} $ & $67.064 \pm 0.846$ & $67.290 \pm 0.088$ \\
$ \tilde m_{K^{*'}} \mathrm{[MeV]}$  &  $1392.2 \pm 57.6$ & $1361.7 \pm 6.3 $  \\
$\tilde \Gamma_{K^{*'}} \mathrm{[MeV]}$ &  $296.67  \pm 160.28$ & $254.62  \pm 17.45$  \\
$\beta$ & $-0.0404\pm 0.0206$  & $-0.0338\pm   0.0023$ \\  
$\lambda_+' \times 10^{3} $ &  $ 25.621 \pm   0.405$ & $ 25.601 \pm   0.277$ \\
$\lambda_+'' \times 10^{3} $ & $ 1.2221 \pm  0.0183$  & $ 1.2150 \pm  0.0090$ \\
\hline 
\hline 
$\chi^2/d.o.f$ & $60.2/68$ &   $28.1/71$  \\
\hline 

\end{tabular}
\caption{
{\it Results for the K$\pi$ form factors parameters from a combined fit to $\tau \rightarrow K \pi \nu_\tau$ and $K_{\ell 3}$.
Note  that $\tilde m_{R}$ and $\tilde \Gamma_R$ are model parameters and do not correspond to the physical 
resonance masses and width. 
}}
\label{tab:fit}
\end{center}
\end{table}

\begin{table}[h!!]
\begin{center}
\begin{tabular}{l|ccccccccc} 
  Parameter  & ln $C$    &  $\lambda_0'$ &  $\tilde m_{K^*}$& $\tilde \Gamma_{K^*}$ & $\tilde m_{K^{*'}}$ & $\tilde \Gamma_{K^{*'}}$ & $\beta$ & $\lambda_+'$& $\lambda_+''$ \\   

ln $C$             & 1 & 0.943 & -0.093 & -0.117 & 0.047 & 0.005 & -0.003 & 0.342 & 0.135 \\
   
$\lambda_0'$       &  -- & 1    &-0.066 & -0.068 & 0.040 & 0.027 &-0.067 & 0.318 & 0.266 \\

$\tilde m_{K^*}$   
                  &   -- & -- & 1  & 0.951 & 0.196 & 0.240 & -0.345 & 0.001 & -0.250 \\
$\tilde \Gamma_{K^*}$  
                 &   -- & -- & --  & 1 &  0.145 & 0.179 &-0.273 & 0.017& -0.160 \\  
$\tilde m_{K^{*'}}$
                 &   -- & -- & --  & -- & 1  & 0.926 & -0.842 & 0.088 & 0.030  \\  
$\tilde{\Gamma}_{K^{*'}}$  
                 &   -- & -- & --  & -- & -- & 1 & -0.917 & 0.088 & 0.030 \\     
$\beta$            
                 &   -- & -- & --  & -- & -- & -- & 1 & -0.128 & -0.018\\  
$\lambda_+'$   
                   &   -- & -- & --  & -- & -- & -- & -- & 1 & 0.735 \\                    
\noalign{\smallskip}\hline
  \end{tabular}
\caption{\label{tab:Corrfit}
{\it Correlations between the parameters of the fit.}}  
  \end{center}
  \end{table}
\begin{table}[h!!]
\begin{center}
\begin{tabular}{lccccc}
\hline
\hline
Integral        &  result &   error & exp & theo  \\
\hline 
$I_{K^0}^\tau$ & 0.50418 & 0.01762 & 0.01689 & 0.00501  \\
$I_{K^0}^{e}$  & 0.15472 & 0.00022 & 0.00022 & 0.00000  \\
$I_{K^0}^\tau/I_{K^0}^{e}$ & 3.25864 & 0.11115 & 0.10634 & 0.03235  \\
\hline
\hline
$I_{K^+}^\tau$ & 0.52387 & 0.01958 & 0.01889 & 0.00515 \\
$I_{K^+}^{e}$  & 0.15909 & 0.00025 & 0.00025 & 0.00000  \\
$I_{K^+}^\tau/I_{K^+}^{e}$ & 3.29282 & 0.12032 & 0.11589& 0.03235 \\
\hline
\hline
\end{tabular}
\caption{{\it Phase space integrals for the charged and neutral modes of $\tau \rightarrow K \pi \nu$ and $K_{e3}$ 
as well as their ratio using the results of the fits to Belle and $K_{\ell3}$ data, see Tab.~\ref{tab:fit}. 
The experimental uncertainty comes from the uncertainties from the fit parameters and the theoretical 
uncertainty comes from the uncertainty of the phase of the form factors in the inelastic region, 
where a large band of 2$\pi$ has been taken, see section 2.3.1. The two uncertainties have been summed 
in quadrature to give the final one.}}
\label{Tab:IkBelle}
\end{center}
\end{table}
\begin{table}[h!!]
\begin{center}
\begin{tabular}{lccccc}
\hline
\hline
Integral        &  result &   error & exp & theo  \\
\hline 
$I_{K^0}^\tau$ & 0.49590 & 0.00820 & 0.00662 & 0.00484  \\
$I_{K^0}^{e}$  & 0.15471 & 0.00015 & 0.00015 & 0.00000  \\
$I_{K^0}^\tau/I_{K^0}^{e}$ & 3.20545 & 0.05060 & 0.03562 & 0.03130  \\
\hline
\hline
$I_{K^+}^\tau$ & 0.51536 & 0.00858 & 0.00631 & 0.00498 \\
$I_{K^+}^{e}$  & 0.15908 & 0.00017 & 0.00017 & 0.00000  \\
$I_{K^+}^\tau/I_{K^+}^{e}$ & 3.23973 & 0.05114 & 0.03635 & 0.03132 \\
\hline
\hline
\end{tabular}
\caption{{\it Phase space integrals for the charged and neutral modes of $\tau \rightarrow K \pi \nu$ and $K_{e3}$ 
as well as their ratio using the results of the fits to the projected 2$^{\mathrm{nd}}$ generation of $B$-factories 
and $K_{\ell3}$ data, see Tab.~\ref{tab:fit}.}}
\label{Tab:IkSuperB}
\end{center}
\end{table}

\subsection{Electromagnetic effects in $\tau \rightarrow K \pi \nu_\tau$}
\label{Sec:em}
While the electromagnetic corrections are known  for $K_{\ell3}$ 
to order $(e^2p^2)$ in ChPT~\cite{Cirigliano:2001mk, Cirigliano:2004pv,Cirigliano:2008wn}, 
they have never been computed in the case of $\tau \rightarrow K \pi \nu_\tau$. 
In this case there are no rigorous methods to compute electromagnetic effects 
over the entire phase space, 
because the kinematics of $\tau$ decays allows 
the hadronic invariant mass squared $s = (p_K + p_\pi)^2$ to extend  
well beyond the chiral regime, all the way to $s = m_\tau^2$. 
While resonance-model calculations are possible~\cite{Cirigliano:2001er, Cirigliano:2002pv}, 
here we will give a first estimate of the long-distance electromagnetic 
corrections to $\tau \to K \pi \nu_\tau$ based on point-like mesons 
and leading Low bremsstrahlung contributions, i.e.  neglecting  structure dependent effects. 
%
%
With these approximations we provide the corrections to both 
differential and total rate for the processes $\tau \rightarrow K \pi \nu_\tau$. 

The leading $O(\alpha)$ long-distance EM corrections arise from one-loop corrections 
to the decay amplitudes and real photon emission.  
Only the one-photon-inclusive decay rate is infrared (IR) finite to $O(\alpha)$. 
Our approach here relies on   the analysis  of   EM corrections to $K \to \pi \ell \bar{\nu}_\ell$   
and $\tau \to \pi \pi \nu_\tau$  presented in  Refs.~\cite{Cirigliano:2001mk, Cirigliano:2004pv} and 
\cite{Cirigliano:2001er, Cirigliano:2002pv}, respectively. 
Adapting the arguments 
presented in Ref. \cite{Cirigliano:2001mk, Cirigliano:2004pv}  we  find  that  long distance EM effects 
in $\tau \to K \pi \nu_\tau$ induce
\footnote{For the two decay modes we adopt this conventions for the particle four-momenta: 
$\tau^-(p_\tau) \rightarrow \pi^-(p_1) K^0(p_2) \nu_\tau(q)$  and 
$\tau^-(p_\tau) \rightarrow  K^-(p_1) \pi^0(p_2) \nu_\tau(q)$. 
The EM  corrections involve the Mandelstam variable $u = (p_\tau - p_1)^2$, 
where $p_\tau$ and $p_1$ denote the four-momentum of the $\tau$ and the charged meson ($K$ or $\pi$) 
in the final state.Moreover,  $m_1^2 = p_1^2$ denotes the mass squared of the charged meson.}:\\

\noindent (i) An overall  correction $g_{\rm rad} (s,u)$  to the differential decay rate, that 
combines the effect of soft real photon emission and the 
universal soft part of one-loop diagrams. The virtual- and real-photon 
corrections are IR divergent and depend on the IR regulator $M_\gamma$, while their sum is finite:
\begin{equation}
g_{\mathrm{rad}}(s,u) \equiv \frac{\alpha}{2\pi} \Gamma_C (u,m_\tau^2,m_1^2,M_\gamma^2) +
 g_{\rm brems} (s,u,m_1^2, m_2^2,M_\gamma^2)~.
\label{eq:grad}
\end{equation}
The expression for  $\Gamma_C (u,m_\tau^2,m_1^2,M_\gamma^2)$ can be found in Ref.~\cite{Cirigliano:2001mk, Cirigliano:2004pv} 
and is reported for completeness in Appendix~\ref{sect:appB}. 
$g_{\rm brems} (s,u,m_1^2,m_2^2,M_\gamma^2)$ encodes  the Bremsstrahlung effects in the leading Low approximation 
and its expression can be found in Ref.~\cite{Cirigliano:2001er, Cirigliano:2002pv} and Appendix~\ref{sect:appC}.\\

\noindent (ii) Shifts  to the form factors: $\bar{f}^{K\pi}_{\pm,0}(s) \to 
\bar{f}^{K\pi}_{\pm,0}(s)  +  \delta \bar{f}^{K\pi}_{\pm,0}(s,u)$. 
These shifts arise already when treating $K$ and $\pi$ as point-like 
as soon as one uses momentum-dependent vertices  for the weak hadronic current. 
$\delta \bar f_\pm(u)$ are given by
\begin{equation}
\delta \bar f_{\pm}^{K^- \pi^0} (u) = 
\frac{\alpha}{4 \pi} \frac{1}{f_+(0)}\left[ \Gamma_1(u,m_\tau^2,m_K^2) \pm \Gamma_2 (u,m_\tau^2,m_K^2) \right] + \dots~, 
\label{eq:deltafK-pi0}
\end{equation}
\begin{equation}
\delta \bar f_{\pm}^{\bar K^0 \pi^-} (u) = 
\frac{\alpha}{4 \pi} \frac{1}{f_+(0)}\left[ \Gamma_2(u,m_\tau^2,m_\pi^2) \pm \Gamma_1 (u,m_\tau^2,m_\pi^2) \right] 
+ \dots~, 
\label{eq:deltafK0pi-}
\end{equation}
The dots denote structure-dependent corrections that are hard to estimate over all 
the phase space. Near threshold, the ChPT expressions in terms of low-energy constants can be found in 
Ref.~\cite{Cirigliano:2001mk, Cirigliano:2004pv}. 
The loop functions  $\Gamma_{1,2} (u,m_\tau^2,m_1^2)$ 
can be found in Ref.~\cite{Cirigliano:2001mk, Cirigliano:2004pv}  and in  Appendix~\ref{sect:appB}.
Finally, in terms of the shifts $\delta \bar{f}_\pm^{K\pi} (u)$,  the corrections to the scalar form factor reads 
$\delta \bar f_{0}^{K \pi} (s,u) \equiv \delta \bar f_+^{K\pi} (u) + s/\Delta_{K\pi}   \   \delta \bar f_-^{K\pi} (u)$.

\begin{figure}[t]
\includegraphics[angle=0,width=0.5\textwidth]{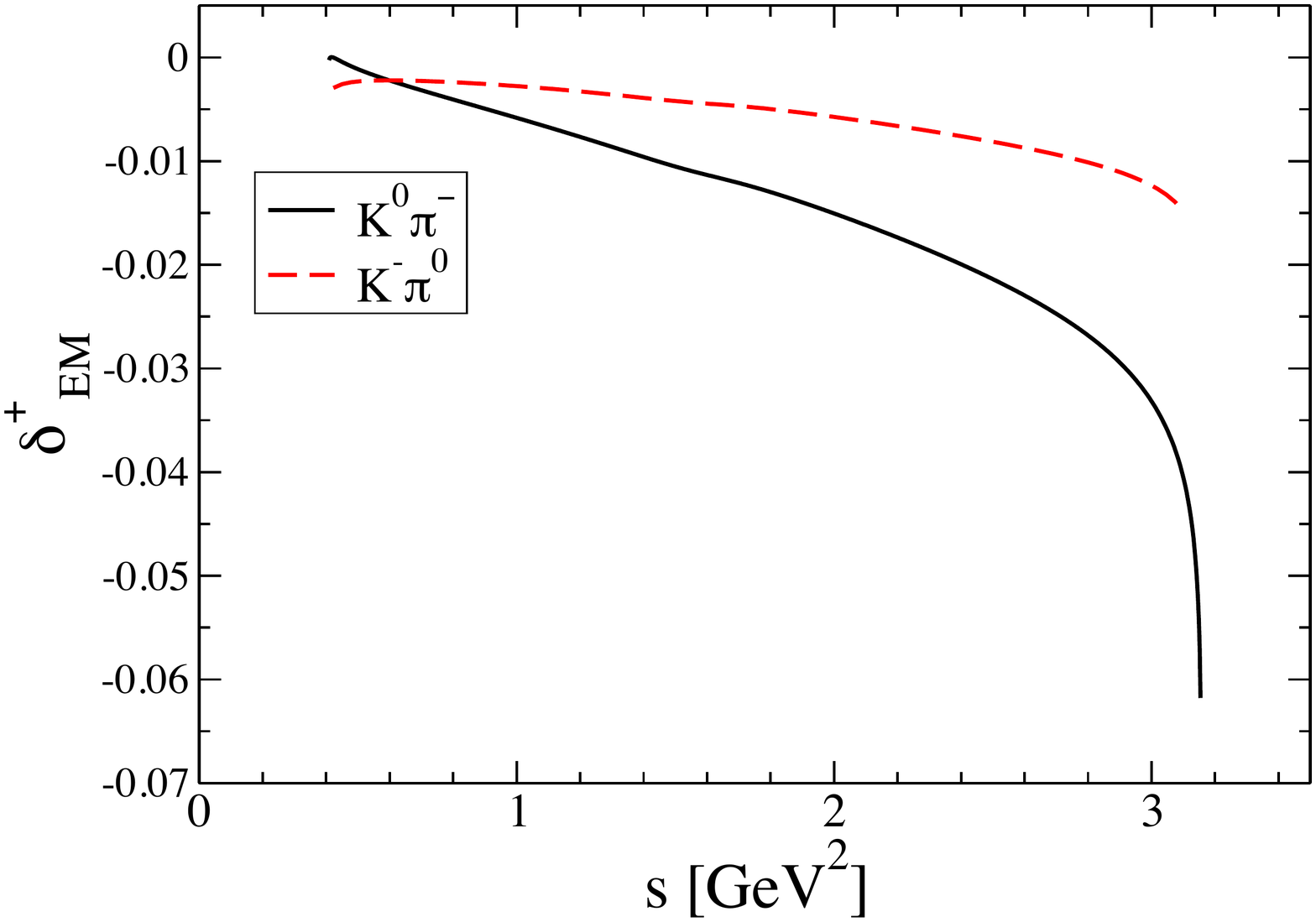}
\includegraphics[angle=0,width=0.5\textwidth]{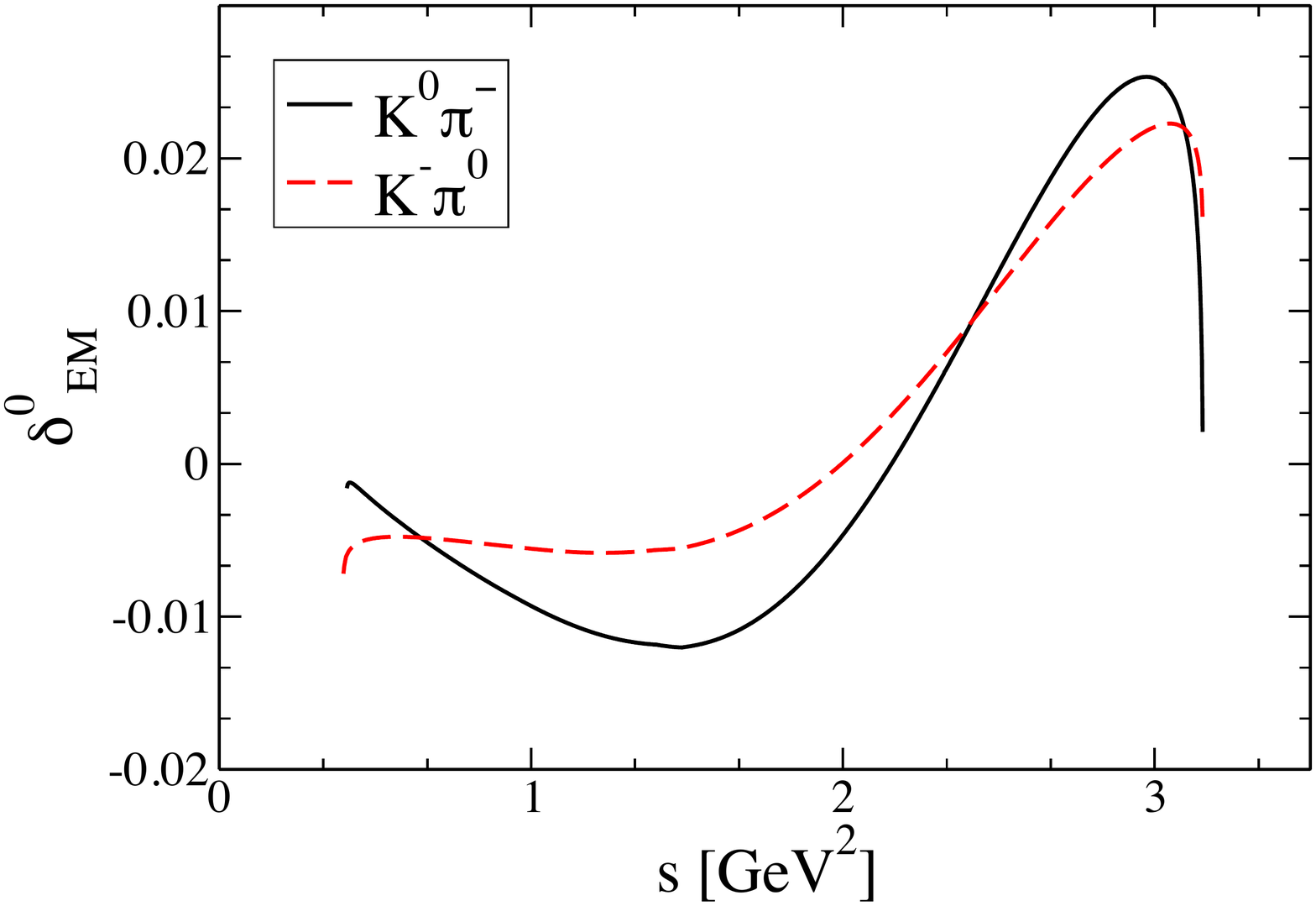}
\caption{\it Correction factors $\delta_{\rm EM}^{+} (s)$  (left panel)
and $\delta_{\rm EM}^{0} (s)$  (right panel) to the vector and scalar contribution to the 
differential decay rates of both $\tau \to K \pi \nu_\tau$ modes.}
\label{fig:GTau}
\end{figure}

With the above prescriptions, and linearizing in the corrections to the form factors, 
we obtain the following expression for the photon-inclusive  double differential rate 
$\tau \rightarrow K \pi \nu_\tau [\gamma]$  decay:
\begin{eqnarray} 
\label{eq:dGsu2} 
\displaystyle\frac{d \Gamma_{\tau  \to K \pi \nu  [\gamma]}}{ds~du} &=& 
\displaystyle\frac{G_F^2 C_{K}^{\tau 2} S_{\rm EW}^\tau |f_+(0)V_{us}|^2}{128 \pi^3 m_\tau^3}
\ \Bigg[
D^{\bar{K} \pi} _+(s,u) \Big( |\bar f_+(s)|^2 + 2 \mathrm{Re} \left[\bar f_{+}(s)\delta \bar f_+^*(u)\right] \big)  \nonumber\\ 
&+& D^{\bar{K} \pi} _0(s,u) \Big( |\bar f_0(s)|^2 + 2 \mathrm{Re} \left[ \bar f_{0}(s) \delta \bar f_0^*(s,u) \right] \Big)  
\\
&+&  D^{\bar{K} \pi} _{+0}(s, u)~\mathrm{Re} \Big[ \bar f_+(s) \bar f^*_0(s) + \bar f_+(s) \delta \bar f^*_0(s,u)+ \delta \bar f_+(u) \bar f^*_0(s) \Big]
\Bigg]  \times  \Bigg[ 1 +  g_{\rm rad} (s,u) \Bigg]~.
\nonumber
\end{eqnarray} 
The expression for the Dalitz plot kinematic  densities  $D_{+,0,+0} (s,u)$ can be found in Appendix~\ref{sect:appA}.
Integrating over the $u$ variable we obtain the EM-corrected distribution in the  $K\pi$ invariant mass:
\begin{eqnarray} 
\label{eq:Gem} 
\displaystyle\frac{d \Gamma_{K \pi [\gamma]}}{ds} \!\!&=&\!\! 
\displaystyle\frac{G_F^2 C_{K}^{2} S_{\rm EW}|f_+(0) \, V_{us}|^2 m_\tau^3 }{96 \pi^3 s \sqrt{s}} 
\left[\left( 1 -\frac{s}{m_\tau^2} \right)^2 
\left(\left(1+\frac{2s}{m^2_\tau}\right)~q^3_{K\pi}(s) |\bar f_+(s)|^2 \left[1+ \delta_{\mathrm{EM}}^+(s) \right] \right. \right. \nonumber \\
& & \left. \left.+ \frac{3 \Delta_{K\pi}^2}{4s}q_{K\pi}(s)\,|\bar f_0(s)|^2 \left[1+ \delta_{\mathrm{EM}}^0(s) \right] \right) + 
\mathrm{Re} \left[ \bar f_+(s) \bar f_0^*(s)\right] \delta_{\mathrm{EM}}^{+0}(s) \right]~,
\end{eqnarray} 
with 
\begin{eqnarray}
\label{eq:deltaem+}
\delta_{\mathrm{EM}}^+ (s)& \equiv &  \frac{ \int_{u_{\rm min}(s)}^{u_{\rm max}(s)}~du~D_{+}(s,u) 
\left(| \bar f_{+}(s)|^2 \, g_{\mathrm{rad}}(s,u) +  2 \mathrm{Re} \left[ \bar f_{+}(s)\delta \bar f_+^*(u)\right]\right)}
{\int_{u_{\rm min}(s)}^{u_{\rm max}(s)}~du~D_{+}(s,u) |\bar f_{+}(s)|^2 }~, \\
\label{eq:deltaem0}
\delta_{\mathrm{EM}}^0 (s) & \equiv &  \frac{ \int_{u_{\rm min}(s)}^{u_{\rm max}(s)}~du~D_{0}(s,u) 
\left(|\bar f_{0}(s)|^2  \, g_{\mathrm{rad}}(s,u) +  2 \mathrm{Re} \left[\bar f_{0}(s)\delta \bar f_0^*(s,u)\right]\right)}
{\int_{u_{\rm min}(s)}^{u_{\rm max}(s)}~du~D_{0}(s,u) |\bar f_{0}(s)|^2 }~, \\
\label{eq:deltaem+0}
\delta_{\mathrm{EM}}^{+0} (s) &\equiv &\frac{3 s \sqrt{s}}{4 m_\tau^6}  \,   \int_{u_{\rm min}(s)}^{u_{\rm max}(s)}~du~D_{+0}(s,u) 
\Big(\mathrm{Re} \left[\bar f_+(s) \bar f^*_0(s) \right]  \, g_{\mathrm{rad}} (s,u) 
\nonumber \\
&+& \  \mathrm{Re} \left[\bar f_+(s) \delta \bar f^*_0(s,u) + \delta \bar f_+(u) \bar f^*_0(s)\right] \Big)~.
\end{eqnarray}
$u_{\rm min,max}(s)$ can be found in the appendix.  
The functions $\delta_{\mathrm{EM}}^{+,0}(s)$ are shown on Fig. \ref{fig:GTau}. 
Further integrating over the $s$  the distribution
(\ref{eq:Gem})  with and without electromagnetic corrections, 
and taking the ratio,  we get $\dEM{K\tau}$.
Assigning an uncertainty of $\sim \alpha/\pi$ to the unknown structure-dependent corrections, 
we get:  
\begin{equation}
\dEM{K^- \tau} = - (0.2 \pm 0.2) \% \qquad \qquad 
\dEM{\bar{K}^0 \tau} = - (0.15 \pm 0.2) \% \
\end{equation}
Note that a comparison between the leading Low approximation and 
the full calculation is performed in Ref.~\cite{Cirigliano:2001er, Cirigliano:2002pv}, and 
it shows that it leads to a comparable  correction to the decay 
rate.  Hence we expect this calculation to give a reasonable estimate for the 
electromagnetic corrections to the  $\tau \rightarrow K \pi \nu_\tau$ total decay rates. 
We have introduced the EM correction factors  $\delta_{\rm EM}^{+,0,+0}(s)$  in the fitting 
procedure and we have found that these corrections do not  affect the determination of the form factors 
at the current level  of precision.  



\subsection{Isospin breaking corrections in $\tau \rightarrow K^-  \pi^0  \nu_\tau$}
\label{Sec:IB}

In order to estimate strong isospin breaking effects, we focus on the dominant  vector form factor. 
We adopt a simple parameterization of  the ratio  $f_+^{K^- \pi^0} (s)/ f_+^{\bar{K}^0 \pi^-} (s)$ 
based on  a single vector meson resonance exchange. 
The ratio  $f_+^{K^- \pi^0} (s)/ f_+^{\bar{K}^0 \pi^-} (s)$ differs from unity because of 
(i) $\pi^0 - \eta$ mixing and (ii)  possible isospin-breaking effects in the coupling of $K^{*-}$ to  $K \pi$. 
To leading order in isospin-breaking, the first effect is independent of $s$, and completely controlled by the 
$\pi^0 - \eta$ mixing angle $\epsilon =   \frac{\sqrt{3}}{4} \frac{m_d - m_u}{ m_s - 1/2 (m_u + m_d)} = 1.16(13) \times 10^{-2}$~\cite{Colangelo:2010et}.
The second effect can be estimated by using  couplings of vector mesons to Goldstone Bosons that involve 
insertions of quark mass matrices, such as those introduced in Ref.~\cite{Cirigliano:2006hb}.  
Requiring that the form factors in the isospin-symmetric limit fall off as $1/s$,  
single vector meson resonance exchange implies the parameterization:
\begin{equation}
f_+^{K^- \pi^0} (s)/ f_+^{\bar{K}^0 \pi^-} (s) = \Big( 1 + \sqrt{3} \, \epsilon \Big) \, 
\left(
1 + \tilde{g}  \frac{m_K^2}{(4 \pi F_\pi)^2} \frac{s}{m_{K^*}^2}  \, \epsilon 
\right)~.
\end{equation}
The only unknown parameter in the above expression is the coupling $\tilde{g} \sim O(1)$, which we vary 
between $-2$ and $+2$. This gives a first rough estimate of the effect of  $s$-dependent isospin breaking
effect, namely $ \tdSU{K^- \pi^0}  =   \pm  0.5 \%$. 
%
%
On the other hand, the constant part due to  $\pi^0 - \eta$ mixing  
is better known and is 100\% correlated with the analogous $K_{\ell 3}$ quantity. 
Putting the two ingredients together, this procedure leads to  $\tdSU{K \pi}  =  (2.9 \pm 0.4_{\rm mixing} \pm 0.5)\% $.
We emphasize that this is a far-from-complete estimate of strong isospin breaking effects, 
and it is only meant to provide a rough estimate of the central value and uncertainty associated  with these effects.

\subsection{Branching ratios}
Using Eq.~(\ref{eq:BrtauKpinu}) we  predict Br($\tau^- \rightarrow  K^-\pi^0 \nu_\tau$) and 
Br($\tau^- \rightarrow \bar K^0\pi^- \nu_\tau$).
In Tab.~\ref{Tab:InputsBrK+}  we summarize the input values used for the predictions. 
We find for the branching ratios 
\begin{equation}
{\rm BR} ( \tau^- \rightarrow \bar K^0 \pi^- \nu_\tau) = (0.8566 \pm 0.0299) \cdot 10^{-2}
\label{eq:BRK0}
\end{equation}
\begin{equation}
{\rm BR} ( \tau^- \rightarrow K^-\pi^0 \nu_\tau) = (0.4707 \pm 0.0181) \cdot 10^{-2}
\label{eq:BRK-}
\end{equation}
with a $100\%$ correlation.  
The error  comes exclusively  from the uncertainty on the $\tau$ phase space integrals. 
In Tab.~\ref{Tab:Brfsfix} results for the 2$^{\rm nd}$ generation of flavour factory with the error budget can be found. 
One sees that the uncertainty coming from the evaluation of the phase space integrals can be reduced by a factor of 
three. Then the uncertainties coming from EM corrections   start to matter. 

\begin{table}[h!!]
\begin{tabular}{lccc|}
\hline \\[-1.5ex]
Parameter          &  Value &   Ref.  \\  [2.pt]
\hline \\[-1.5ex]
BR$(K^{\pm}_{e3})$ & 0.05078(31)              & \cite{Antonelli:2010yf} \\
$\tau_{K^{\pm}}$   & (12.384 $\pm$ 0.015)ns   & \cite{Antonelli:2010yf} \\
$\tau_\tau$        & (290.6 $\pm$ 1.0) fs     & \cite{Beringer:1900zz} \\
\\[-2.5ex]
\hline \\[-1.5ex]
$S_{EW}^K$         & 1.0232 $\pm$ 0.0003      &  \cite{Marciano:1993sh} \\
$S_{EW}^\tau$      & 1.0201 $\pm$ 0.0003      &  \cite{Erler:2002mv} \\
$\dEM{K \ell} (\%)$& 0.050  $\pm$ 0.125       &  \cite{Cirigliano:2008wn} \\
$\dEM{K^- \tau}$       &    -(0.2 $\pm$ 0.2) \%                      & Sec.~\ref{Sec:em} \\
$\dSU{K \pi}$      & 0.029 $\pm$ 0.004        & \cite{Antonelli:2010yf,Kastner:2008ch} \\ 
$\tdSU{K\pi}$       & 0.029 $\pm$ 0.006        & Sec.~\ref{Sec:IB} \\
$I_{K^+}^\tau/I_{K^+}^{e}$ & 10.32059 $\pm$ 0.48240 & Tab. \ref{Tab:IkBelle}\\
\\[-2.5ex]
\hline
\end{tabular}
\begin{tabular}{lccc}
\hline \\[-1.5ex]
Parameter          &  Value &   Ref.  \\ [2.pt]
\hline \\[-1.5ex]
BR$(K_{L e3})$     &  0.4056(9)               & \cite{Antonelli:2010yf} \\
$\tau_{K_{L}}$     & (51.16 $\pm$ 0.21)ns     & \cite{Antonelli:2010yf} \\
$\tau_\tau$        & (290.6 $\pm$ 1.0) fs     & \cite{Beringer:1900zz} \\
\\[-2.5ex]
\hline \\[-1.5ex]
$S_{EW}^K$         & 1.0232 $\pm$ 0.0003      &  \cite{Marciano:1993sh} \\
$S_{EW}^\tau$      & 1.0201 $\pm$ 0.0003      &  \cite{Erler:2002mv} \\
$\dEM{K \ell} (\%)$& 0.495  $\pm$ 0.110       &  \cite{Cirigliano:2008wn} \\
$\dEM{\bar{K} ^0\tau}$       &                -(0.15 $\pm$ 0.2) \%             & Sec.~\ref{Sec:em} \\
$\dSU{K \pi}$      & 0                        & \cite{Antonelli:2010yf,Kastner:2008ch} \\ 
$\tdSU{K \pi}$       & 0                        & Sec.~\ref{Sec:IB} \\
$I_{K^+}^\tau/I_{K^+}^{e}$ & 10.21432 $\pm$ 0.43058 & Tab. \ref{Tab:IkBelle}\\
\\[-2.5ex]
\hline
\end{tabular}
\caption{\it Input used to compute $\tau^- \rightarrow K^-\pi^0 \nu_\tau$ 
and  $\tau^- \rightarrow \bar K^0\pi^- \nu_\tau$ 
branching ratios.}
\label{Tab:InputsBrK+}
\end{table}

\begin{table}[h!!]
\begin{center}
\begin{tabular}{lccccccccc}
\hline
\hline
Mode & BR & $\%$ err & BR($K_{e3}$) & $\tau_K$ & $\tau_\tau$  & $I_{K}^\tau/I_{K}^{e}$ & 
EM  & SU(2) \\
\hline 
$\tau^- \rightarrow \bar K^0 \pi^- \nu_\tau$ & 0.8427 $\pm$ 0.0122 & 1.45 & 0.22 & 0.41 & 0.34 & 1.24 & 0.46 & 0 \\
$\tau^- \rightarrow K^-\pi^0 \nu_\tau$ & 0.4631 $\pm$ 0.0079 & 1.71 & 0.06 & 0.12 & 0.34 & 1.25 & 0.47 & 1.00 \\
\hline
\hline
\end{tabular}
\caption{{\it Prediction for the $\tau^- \rightarrow K^-\pi^0 \nu_\tau$ branching fractions in $\%$ from the $K_{e3}$ 
branching ratio using the 2$^{\mathrm{nd}}$ generation of B factory projected results for the phase space integrals, see Tab.~\ref{Tab:IkSuperB}. 
The different sources of uncertainty are given. They have been summed in quadrature to give the total  one.}}
\label{Tab:Brfsfix}
\end{center}
\end{table}

\section{Implications for the inclusive determination of   $V_{us}$}
\label{sect:4}

\begin{figure}[t]
\begin{center}
\includegraphics[width=0.8\textwidth]{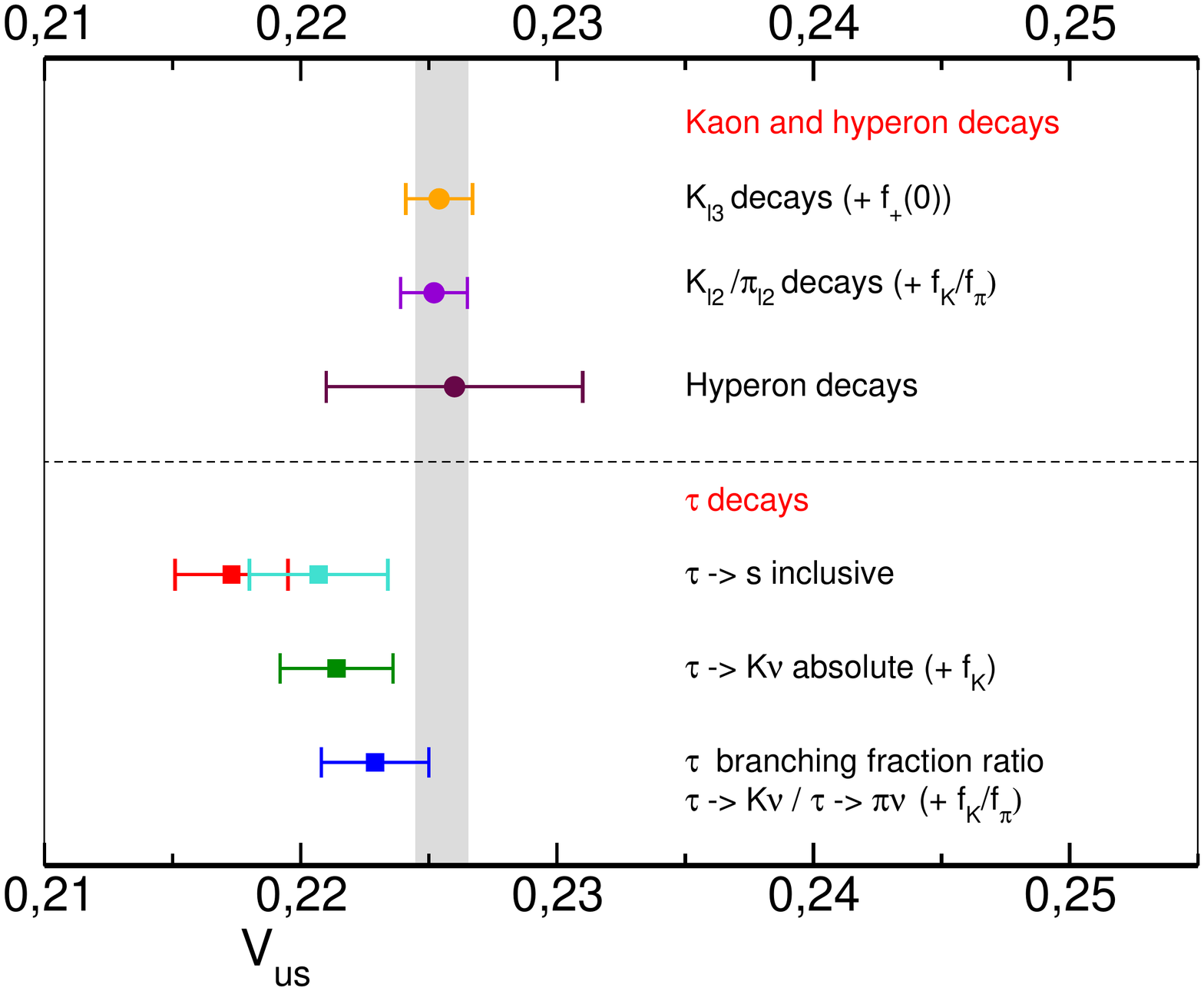}
\caption{\it 
Determination of $|V_{us}|$ from semileptonic, leptonic kaon decays~\cite{Antonelli:2010yf}, hyperon decays~\cite{Mateu:2005wi} 
and inclusive and exclusive $\tau$ decays~\cite{Amhis:2012bh}. 
The errors bars correspond to the determination from exclusive $\tau$ decays (blue),  the inclusive hadronic 
$\tau$ decays (red), and our prediction (cyan).  
The grey band displays the value of $|V_{us}|$ assuming unitarity of the first row of the CKM matrix.
}
\label{fig:Vus}
\end{center}
\end{figure}

The most precise determination of $|V_{us}|$ from $\tau$ decays comes from the measurements of inclusive 
$|\Delta S| =0$ and $|\Delta S| = 1$ tau decay widths. Indeed one can build the theoretical quantity 
\begin{equation}
\delta R_{\tau,th} = \frac{R_{\tau,NS}}{|V_{ud}|^2}- \frac{R_{\tau,S}}{|V_{us}|^2}~,
\label{Eq:DeltaSU(3)}
\end{equation}
where $R_{\tau}$ is defined as
\begin{equation}
R_{\tau}=\frac{\Gamma [\tau \to {\rm hadrons}~\nu_\tau]}{\Gamma[\tau \to \bar \nu_e e \nu_\tau] }~.
\end{equation} 
This quantity vanishes in the $SU(3)$ limit and can be precisely determined within QCD combining perturbative QCD and 
low energy data~\cite{Gamiz:2002nu, Gamiz:2004ar, Maltman:2006st}. 
Hence, we can extract $V_{us}$ from Eq.~(\ref{Eq:DeltaSU(3)}) using the theoretical estimate of  $\delta R_{\tau,th}$ and 
the precise measurements of non-strange ($R_{\tau,NS}$) and strange ($R_{\tau,S}$) inclusive decays, 
and $|V_{ud}|$. 
Following Ref.~\cite{Amhis:2012bh}, we take $\delta R_{\tau,th} = 0.240 \pm 0.032$, with
a systematic error on $|V_{us}|$ that lies between the two more recent estimates~\cite{Gamiz:2007qs,Maltman:2010hb}.  
We use $|V_{ud}| =0.97425 \pm 0.00022$ from the superallowed $0^+ \to 0^+$ nuclear $\beta$ decays~\cite{Hardy:2008gy}. 
Using the HFAG Early 2012 averages from the $\tau$ branching fractions reported in Table~\ref{Tab:TauBrstrange} 
together with their reported statistical correlations and replacing the $\tau \to K \nu_\tau$ and $\tau\rightarrow K\pi \nu_\tau$ results  
by our prediction in Eq.~(\ref{eq:BRtauKnu}) and Eqs.~(\ref{eq:BRK0}), (\ref{eq:BRK-}), we obtain~\cite{Amhis:2012bh}
\begin{equation}
{\rm BR}_{\tau,S} \equiv {\rm BR}(\tau \rightarrow X_s^- \nu_\tau)= (2.9648 \pm 0.0656)\cdot 10^{-2}~.
\label{eq:tauXs}
\end{equation}
With this estimate and ${\rm BR}_{\tau,NS} = (61.85 \pm 0.11)\%$ and ${\rm BR}_{\tau,e} = (17.839 \pm 0.028)\%$~\cite{Amhis:2012bh}, we get
\begin{equation}
|V_{us}|=0.2207 \pm 0.0027~.
\end{equation}
We summarize in Fig.~\ref{fig:Vus} the different extractions of $|V_{us}|$ from semileptonic and leptonic kaon decays, hyperon 
and $\tau$ decays. Our prediction shifts the inclusive determination of $|V_{us}|$ towards the exclusive one by $\sim 1\sigma$.

\section{Conclusion}
\label{sect:5}

The experimental precision of data on leptonic and semileptonic kaon decays matched by 
sub-percent theoretical calculations allowed the most accurate determination of $V_{us}$~\cite{Antonelli:2010yf}. 
Assuming lepton universality, we use the same data in combination with the measurement of the $K \pi$ 
invariant mass distribution in the $\tau \rightarrow K \pi \nu$ decay and a dispersive parametrization for the 
form factors~\cite{Bernard:2011ae, Bernard2013}
to obtain a precise prediction for about 68\%
of the total strange width. A first evaluation of electromagnetic and SU(2) breaking effects has been derived to this purpose. We find:
$$
\mathrm{BR}(\tau^- \to K^- \nu_\tau) = (0.713 \pm 0.003)\times 10^{-2}~,
$$
$$
{\rm BR} ( \tau^- \rightarrow \bar K^0 \pi^- \nu_\tau) = (0.8566 \pm 0.0299) \cdot 10^{-2}~,
$$
$$
{\rm BR} ( \tau^- \rightarrow K^-\pi^0 \nu_\tau) = (0.4707 \pm 0.0181) \cdot 10^{-2}~,
$$
$$
B_3 \equiv \mathrm{BR}(\tau \to K \nu_\tau)+ {\rm BR} ( \tau^- \rightarrow \bar K^0 \pi^- \nu_\tau) +{\rm BR} ( \tau^- \rightarrow K^-\pi^0 \nu_\tau) = ( 2.040 \pm 0.048)\cdot 10^{-2}~.
$$
$B_3$ is $1.7 \sigma$ higher with respect to the world average measured value. 
In addition we obtain a determination of $V_{us}$ from inclusive tau decays using the above prediction for the branching ratios and 
the world average values for the rest of tau branching fractions. We find:
$$
|V_{us}|=0.2207 \pm 0.0027~,
$$
and for the unitarity of the CKM quark mixing matrix as applied to the first row,  we obtain: 
$$1-|V_{us}|^2 - |V_{ud}|^2 = 0.0021 \pm 0.0013~(-1.6\sigma)~.$$ 
Finally, we have shown that  measurements of the $K \pi$ invariant  
mass distribution at a second generation B factory with integrated luminosity 
of 40 ab$^{-1}$ would
reduce the uncertainty in the  $\tau \to K \pi \nu_\tau$ BRs by a  
factor of three,
and therefore further reduce the error on $V_{us}$. 
\appendix
\section{Kinematic densities}
\label{sect:appA}

The differential decay rate Eq.~(\ref{eq:dGsu2}) involves the kinematic functions
\begin{eqnarray}
D^{\bar{K} \pi}_+(s,u) \! \!& =& \! \! \frac{m_\tau^2}{2} (m_\tau^2-s) + 2 m_1^2 m_2^2 - 2u~(m_\tau^2-s+m_1^2+m_2^2)+ 2u^2 \nonumber \\
& & -\frac{\Delta_{21}}{s} m_\tau^2 (2u+s-m_\tau^2-2m_2^2)+\frac{\Delta_{21}^2}{s^2} \frac{m_\tau^2}{2} (m_\tau^2-s)~, \\
D^{\bar{K} \pi}_0(s,u) \! \!& =& \! \! \frac{\Delta_{21}^2 m_\tau^4}{2s^2} \left( 1- \frac{s}{m_\tau^2}\right)~, \\
D^{\bar{K} \pi}_{+0}(s,u) \! \!& =& \! \! \frac{\Delta_{21} m_\tau^2}{s} \left( 2u+s-m_\tau^2 -2 m_2^2 
-\frac{\Delta_{21}}{s} \left(m_\tau^2 -s \right) \right)~,
\end{eqnarray}
with $\Delta_{21} = m_2^2 - m_1^2$.
The above expressions are valid for  both decay modes, with the following conventions for the particle four-momenta: 
$\tau^-(p_\tau) \rightarrow \pi^-(p_1) K^0(p_2) \nu_\tau(q)$  and 
$\tau^-(p_\tau) \rightarrow  K^-(p_1) \pi^0(p_2) \nu_\tau(q)$. 
The Mandelstam variable $u = (p_\tau - p_1)^2$, 
where $p_\tau$ and $p_1$ denote the four-momentum of the $\tau$ and the charged meson ($K$ or $\pi$) 
in the final state. Moreover,  $m_1^2 = p_1^2$ denotes the mass squared of the charged meson.

\section{Loop functions}
\label{sect:appB}
We now give expressions for the loop functions characterizing virtual photon 
corrections. 
We denoted by $M$ the charged meson mass,  by $m_\ell \to m_\tau$ the 
charged lepton mass, and by $M_\gamma$ the photon mass used as infrared regulator. 
In order to express the loop functions $\Gamma_{1,2,C}$  in a compact way, 
it is useful to define the following intermediate variables:
\begin{equation}
 R  =  \frac{m_{\ell}^2}{M^2}  , \ \ 
 Y = 1 + R - \frac{v}{M^2}  , \ \     
 X = \frac{Y - \sqrt{Y^2 - 4 R}}{2 \sqrt{R}} . 
\end{equation} 
In terms of such variables,  of the dilogarithm 
\begin{equation}
Li_2 (x) = - \int_{0}^{1} \frac{dt}{t} \log (1 - x t)   , 
\end{equation}
and the auxiliary functions
\begin{eqnarray}
{\cal C} (v,m_\ell^2,M^2)  &=&   \frac{1}{m_\ell M} \frac{X}{1 - 
X^2} \, 
\Big[  - \frac{1}{2} \log^2 X + 2 \log X \log (1 - X^2) - 
\frac{\pi^2}{6} + \frac{1}{8} \log^2 R 
\nonumber \\
&+&    Li_2 (X^2) + Li_2  \left(1 - \frac{X}{\sqrt{R}}\right) + 
Li_2 (1 - X \sqrt{R})  \Big]
\\
{\cal F}(v,m_{\ell}^2,M^2)\,&=&\frac{2}{\sqrt{R}}\,\frac{X}{1-X^2}\,\ln X ~, 
\end{eqnarray}
%
%
we have:
\begin{equation}
\Gamma_C (v,m_{\ell}^2,M^2 ; M_\gamma^2) = 
2 M^2 Y \, {\cal C} (v,m_\ell^2,M^2)  
+ 2  \log \frac{M m_\ell}{M_\gamma^2} 
\bigg(1 + \frac{X Y 
\log X}{\sqrt{R} 
(1 - X^2)}  \bigg) 
\end{equation}
and 
\begin{eqnarray}
\Gamma_1(v,m_{\ell}^2,M^2) &=& \frac{1}{2} \Big[ -\,\ln R\,+\,
(4-3Y){\cal F}(v,m_{\ell}^2,M^2) \Big]
\nonumber\\
\Gamma_2(v,m_{\ell}^2,M^2) &=& \frac{1}{2} \Big(1-\frac{m_{\ell}^2}{v}\Big) 
\Big[ - {\cal F}(v,m_{\ell}^2,M^2)(1-R)
+ \ln R \Big]  -
\frac{1}{2}(3-Y){\cal F}(v,m_{\ell}^2,M^2)\,.   \ \ \ \ 
\end{eqnarray}

\section{Real photon emission}
\label{sect:appC}

In Refs.~\cite{Cirigliano:2001er,Cirigliano:2002pv}  the function $g_{\rm brems} (s,u,m_1^2,m_2^2,M_\gamma^2)$ 
is denoted  by $g_{\rm brems} (s,u,M_\gamma^2)$, omitting the dependence on the meson masses $m_{1,2}^2$.
We report here the full expressions for completeness. The Bremsstrahlung function is given by:
\begin{eqnarray}
g_{\rm brems} (s,u,M_\gamma) &=& \displaystyle\frac{\alpha}{\pi}
[J_{11}(s,u,M_\gamma)+J_{20}(s,u,M_\gamma)+J_{02}(s,u,M_\gamma)] ~. 
\\
J_{11}(s,u,M_\gamma) &=& \log\left(\frac{2 x_+ (s,u) \bar{\gamma}}{M_\gamma}\right)
\displaystyle\frac{1}{\bar{\beta}}\log\left(\frac{1+\bar{\beta}}
{1-\bar{\beta}}\right) \\
&+& \displaystyle\frac{1}{\bar{\beta}}\left\{ Li_2(1/Y_2) - Li_2(Y_1)+
\log^2(-1/Y_2)/4 - \log^2(-1/Y_1)/4 \right\} \\
J_{20}(s,u,M_\gamma) &=& \log\left(\frac{M_\gamma (m_\tau^2 - s)}
{m_\tau x_+ (s,u)}\right) \\
J_{02}(s,u,M_\gamma) &=& \log\left(\frac{M_\gamma (m_\tau^2 + m_2^2 -s-u)}
{m_1  x_+ (s,u)}\right) \label{eq:Jmn}~.
\end{eqnarray}
The auxiliary variables are: 
\begin{eqnarray}
x_{\pm} (s,u) & = & 
\frac{1}{2 \, m_1^2} \Bigg[ 2 \, m_1^2 \, (m_{\tau}^2 + s) - 
(s + m_1^2 - M_2^2) \, 
 (m_{\tau}^2 + m_1^2 - u) 
 \nonumber \\ 
 & \pm & 
 \sqrt{ \lambda(s, m_1^2, m_2^2) \,  
\lambda(u, m_1^2, m_{\tau}^2) } \Bigg] 
\\
Y_{1,2} &=& \frac{1 - 2 \bar{\alpha} \pm \sqrt{(1 - 2 \bar{\alpha} )^2 - (1 - 
\bar{\beta}^2 )} }{1  + \bar{\beta}} \\
  &  & \nonumber \\
\bar{\alpha} &=& \displaystyle\frac{(m_\tau^2 - s)  
(m_\tau^2 + m_2^2 - s - u)}{
  (m_1^2 + m_\tau^2 - u)} \cdot \frac{\lambda (u, m_1^2, m_\tau^2)}
{2 \, \bar{\delta} }
\\
  &  & \nonumber \\
\bar{\beta} &=& - \frac{\sqrt{ \lambda (u, m_1^2, m_\tau^2) 
   }  }{m_1^2 + m_\tau^2 - u   }   \\
  &  & \nonumber \\
\bar{\gamma} &=& \frac{ \sqrt{  \lambda (u, m_1^2, m_\tau^2)    }}{
2 \, \sqrt{ \bar{\delta}} }      \\
  &  & \nonumber \\
\bar{\delta} &=& - m_2^4 m_\tau^2 + m_1^2 (m_\tau^2 - s) (m_2^2 - u) - 
s  u ( - m_\tau^2 + s + u) \nonumber \\
& &  \qquad  + m_2^2 (- m_\tau^4 + s  u + m_\tau^2 s 
+ m_\tau^2 u)~, 
\end{eqnarray}
with $\lambda(x,y,z) = x^2 + y^2 + z^2 - 2 \,  (x y + x z + y z )$.


\vspace{9mm}
{\Large\bf Acknowledgements\/}\\\\
VC and EP acknowledge support from DOE Office of Nuclear Physics and the LDRD program at Los Alamos National Laboratory.

\bibliographystyle{doiplain}
\bibliography{tau}


\end{document}